\documentstyle[epsfig]{aipproc}

\begin{document}

\title{Continuum Spectra of Quasar Accretion Disk Models}
 
\author{Eric Agol$^*$, Ivan Hubeny$^{\dagger}$, and
Omer Blaes$^{\diamond}$}
\address{$^*$Physics and Astronomy Department, Johns Hopkins University,
Baltimore, MD 21218\\
$^{\dagger}$NASA Goddard Space Flight Center, Code 681, Greenbelt, MD 20771\\
$^{\diamond}$Physics Department, University of California, Santa Barbara,
CA 93106}

\maketitle

\begin{abstract}

We have calculated the spectrum and polarization of a standard thin accretion
disk with parameters appropriate for a bright quasar.  This model improves 
upon previous work by including ultraviolet metal line opacities, assumed for
now to be in LTE.  Though not yet fully self-consistent, our calculations 
demonstrate 
that metal lines can change the spectral slope, reduce the polarization, and 
reduce the Lyman edge feature in accretion disk spectra.  Some observational
differences between quasar spectra and accretion disk models might be
reconciled with the inclusion of metal lines.

\end{abstract}


The prodigious luminosities and compact sizes of quasars led theorists to
a model in which gas falls into a supermassive black hole through a
geometrically thin, optically thick accretion disk.  If quasars contain
a hole of about $10^{8-9} M_\odot$ which is accreting at near the
Eddington limit, then their spectra should peak in the ultraviolet, as
observed \cite{shi78}. However, so far this paradigm has failed to explain 
many other details of quasar spectra:  the calculated spectra are too narrow 
in range of frequency (an extra power law is needed to fit the spectrum
\cite{sun89}); have too strong a feature at the Lyman edge \cite{ant89};  
and have too large polarization at the wrong polarization angle \cite{sto84}.  
Part of the reason for the failure might be due to oversimplified disk 
spectra calculations, which until now have neglected the effects of metal 
line opacity.  We present here a preliminary calculation including
metal line opacities;  the calculation, however, is not fully
self-consistent, but simply illustrates that the opacity of metal lines
can play an important role in shaping accretion disk spectra.

\begin{figure} 
\centerline{\epsfig{file=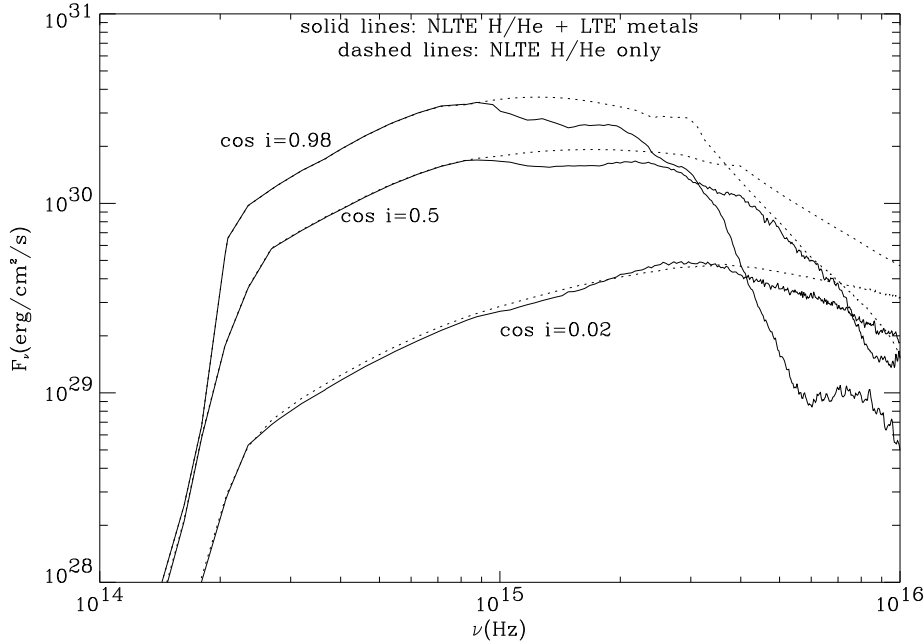,width=5in}}
\vspace{10pt}
\caption{Comparison of the flux ($F_\nu$) at three different viewing angles.
}\label{fig1}
\end{figure}

We have computed the spectrum and polarization of a 
disk with the following parameters:  mass of black
hole ${\rm M_{BH}=2\times 10^{9} M_\odot}$, accretion rate 
${\rm \dot{M}=1 M_\odot/}$year, and
spin of black hole $a=0.998 {\rm M_{BH}}$.  This corresponds to a
luminosity of 0.072 ${\rm L_{Edd}}$.  
We computed the vertical
structure at each radius using TLUSDISK \cite{hub97}. Using
this structure, we then computed the spectrum and polarization using
the code SYNSPEC \cite{hub94} combined with the polarization code of 
Blaes and Agol \cite{bla96}. Finally, we convolved the spectra
from each radius with a relativistic transfer function \cite{ago97}.
The vertical structure (i.e. temperature, density, flux) is computed taking 
into account the continuum opacity of hydrogen and helium in non-LTE, and 
then this structure is used to compute the opacity due to metal lines
assumed to be in LTE at the calculated temperature and density. 
We assume a 30 km/s Doppler width for all lines. We take the outer edge of the 
disk to be $50 {\rm GM_{BH}/c^2}$.  We take into account
departures from LTE for the 9 levels of HI, 14 levels of HeI, and
14 levels of HeII.  So far, we have only included metal lines between 
$200-3000{\rm\AA}$, assuming solar abundance.  We have not yet 
taken into account the effects of bound-bound transitions on the hydrogen 
and helium number densities.  Because the atmosphere structure calculation 
did not include metal lines in the radiative equilibrium condition, the spectrum
is not fully self-consistent.  Hence there is an artificial reduction in flux.
The spectra for different inclination angles are presented in figure~\ref{fig1},
comparing spectra models with and without metal lines.
The flux falls more steeply in the UV when metals are included; above 
$10^{14.8}$~Hz for the face-on disk, $f_\nu\propto \nu^{-0.3}$ approximately.
This may change when the calculation is done self-consistently (i.e. when
flux is conserved).  The models are identical below $10^{15}$~Hz since metal
lines are not included in that wavelength range.  The spectral slope at smaller 
frequencies changes due to the small outer cutoff radius we used.

\begin{figure} 
\centerline{\epsfig{file=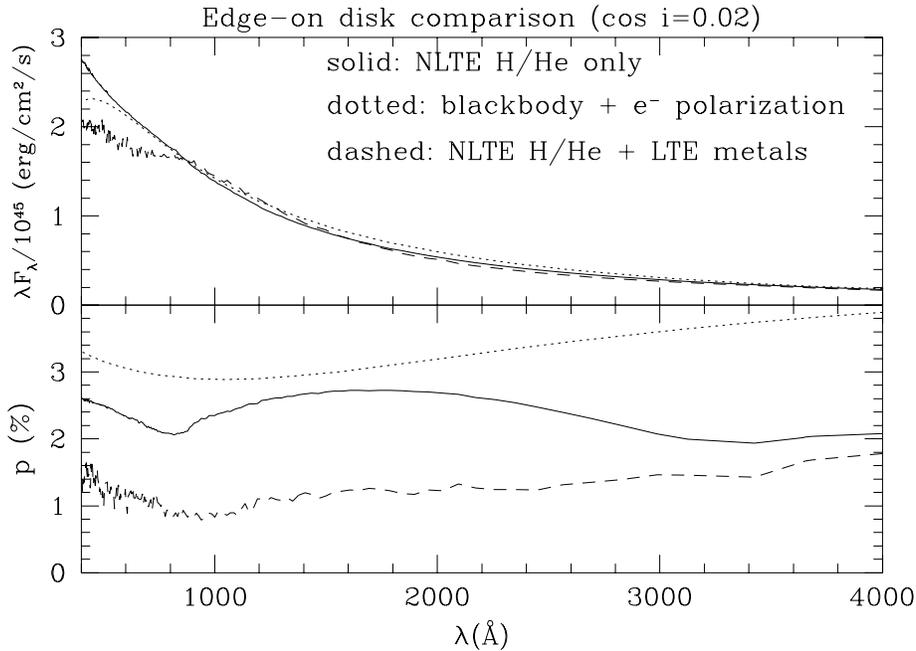,width=5in,height=3.52in}}
\vspace{10pt}
\caption{Comparison of the flux ($\lambda F_\lambda$) and percent polarization 
for the model viewed
nearly edge-on.  The polarization angle is not shown since it stays
roughly constant within 5$^\circ$ for all three models in this wavelength
range. Note that for this viewing angle, the Lyman edge is so highly smeared
that it is not visible, and the continuum is quite hard.}\label{fig2}
\end{figure}

The ultraviolet
polarization is decreased significantly by the line opacity (figure 
\ref{fig2}).
For this nearly edge-on disk, the polarization
is between 1.2-1.8\% in the observable region.  This is reduced from the
maximum value of 3-4\% assuming pure electron scattering with relativistic
effects included.  If quasars are preferentially seen closer to face-on,
then their ultraviolet polarization will be 
even lower, and the small observed polarization could be due to scattering off
of material at larger radii.  

\begin{figure} 
\centerline{\epsfig{file=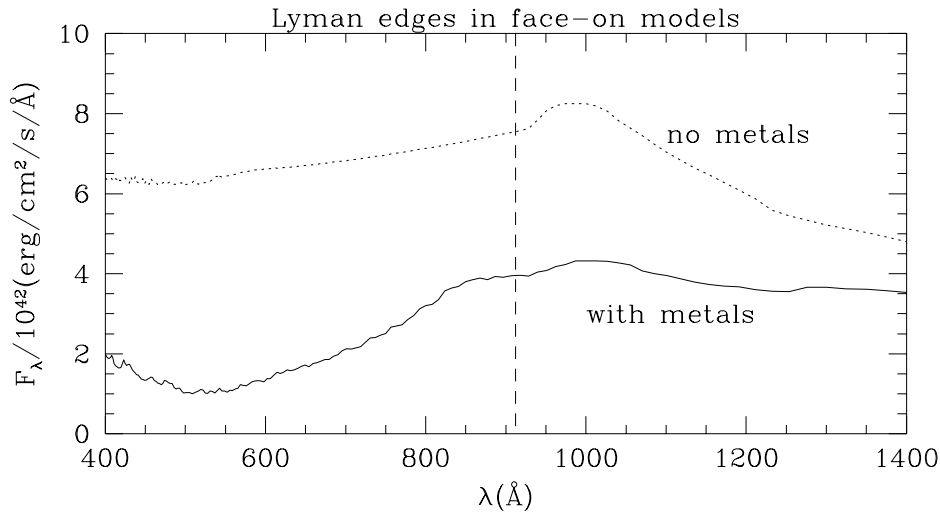,width=5in}}
\vspace{10pt}
\caption{Flux versus wavelength near the Lyman edge
for a face-on ($\cos i=0.98$) disk.  The Lyman edge feature is in emission in 
the inner parts of the disk (which are strongly redshifted), and absorption 
in the outer parts (which are less redshifted), which causes the bump
redward of the Lyman edge.  When metal lines are included, the photosphere
redward of the Lyman edge is brought closer to the surface where the
temperature is higher/lower for an edge in emission/absorption, reducing
the contrast across the Lyman edge.}\label{fig3}
\end{figure}

Inclusion of metal opacity reduces the bump near the Lyman edge which is
present in the H/He continnum-only face-on disk model (figure \ref{fig3}).
This may be why it is difficult to find Lyman edge features in quasars. 
The main contributors to the metal line opacity in this region are
Fe, Ni, Mn, and S.  There is a 
broader dip blueward of the Lyman edge due to metal lines, mostly Fe.  Note 
that the flux is different in the two cases because we haven't included the 
lines in the atmosphere structure calculation, so the total flux is not 
constant.

In summary, metal line opacities can reduce the polarization, change the
spectral shape, and reduce the Lyman edge jump in accretion disk model
spectra for quasars.  We need to see whether these effects still
occur in models which include the lines in calculating the disk structure.
To make a better comparison of the models with and without metals, we will 
construct line-blanketed disk structure models including lines from a wider
range of wavelengths and bound-bound transitions for H and He, and we will 
include the 
contribution of the disk at larger radii.  The metal line opacity will change
the temperature equilibrium and the radiative acceleration, and thus will
be important for calculating the disk structure, which will in turn affect
the continuum shape.


\end{document}